# Measured LWC-Specific Fog Attenuation and Frequency Scaling in Low-THz Channels

Jiabiao Zhao, Xiaoxiang Li, Kefeng Huang, Yapeng Ge, Hanchen Liu, Jie Yang, Weidong Hu, Houjun Sun and Jianjun Ma, *Member*, *IEEE*

***Abstract*—Fog can reduce the link margin of terahertz (THz) wireless systems. Earlier channel measurements mainly relied on visibility, and almost no liquid water content (LWC) referenced attenuation coefficients have been reported. This letter reports controlled fog measurements at low-THz frequencies (120, 140, and 160 GHz) over a 22 m channel. LWC is retrieved from a time-aligned droplet size distribution (DSD) and paired with the fog-induced attenuation to obtain the relationship between attenuation and LWC at each frequency. The comparison with ITU-R P.840 is posed as an errors-in-variables problem. This separates the absolute coefficient from its frequency dependence - how the coefficient grows with frequency. Expressed as a power law of frequency, the measured exponent is 1.347, matching the value of 1.343 implied by P.840 at 20 °C. The results provide reference data and a compact scaling law for fog link budgeting at low-THz frequencies.**

## I. INTRODUCTION

THE terahertz (THz) band, conventionally 0.1-10 THz, provides broad spectral resources for high-capacity short-range links, backhaul, sensing, imaging, and high-resolution radar [1][2]. Its lower reaches, the low-THz band, are the first to approach deployment. International Telecommunication Union Recommendation (ITU-R) F.2172-0 has harmonized fixed-service channel arrangements across 130-174.8 GHz for wireless backhaul and fronthaul [3], where link budgets must account for atmospheric attenuation. Beyond gaseous absorption, suspended liquid droplets introduce absorption and scattering, so fog can erode the link margin even in the absence of rain [4],[5] .

ITU-R P.840 describes fog attenuation as $\gamma_f = K_l\ M$, where $M$ denotes the LWC and $K_l$, in (dB/km)/(g/m$^3$), depends on frequency and temperature through the dielectric response of liquid water in the Rayleigh limit for droplets much smaller than the wavelength [6][7]. P.840 supplies $K_l$ only through the double-Debye formulation. Unlike rain, whose frequency dependence is standardized in closed-form power-law form [8], fog attenuation is not expressed as a comparable power law in any THz band. The descriptors that drive these models are not interchangeable. LWC fixes the total liquid mass per unit volume but not its partition among droplet sizes, the DSD carries that partition, and visibility characterizes optical extinction alone. None of the three determines the others uniquely.

Controlled fog experiments at millimeter-wave and THz frequencies have characterized droplet-induced loss through various descriptors, including visibility, optical extinction, LWC, and DSD [9][10][11]. These studies confirmed measurable attenuation from suspended droplets, but cross-study comparison remains difficult because frequency, path length, fog-generation method, and attenuation definitions differ substantially. Moreover, fog microphysics evolves during formation and dissipation, so establishing the droplet-attenuation relationship reliably requires electromagnetic and microphysical measurements synchronized in time [12][13].

Fog attenuation near 140 GHz, one of the most extensively investigated frequencies for THz communications [2], was examined in early work. But those measurements relied chiefly on visibility rather than on directly retrieved LWC or DSD [5]. More recently, joint propagation and microphysical measurements at 220 and 320 GHz demonstrated DSD-driven dynamic modeling of THz fog channels and confirmed that extinction below 1 THz is well described by Rayleigh, absorption-dominated models [14]. However, these studies focused on higher frequencies and dynamic channel modeling rather than extracting repeated attenuation-LWC coefficients across the low-THz range. A systematic comparison with ITU-R P.840 that separates absolute coefficient agreement from frequency-scaling validation has also not been reported.

This letter addresses these gaps through controlled fog measurements. Time-aligned DSD measurements are conducted to retrieve the LWC and yield the attenuation-LWC relationship at each frequency from several independent runs. The comparison with ITU-R P.840 is formulated as an errors-in-variables problem, to separate the uncertainty in the absolute coefficient from the frequency-dependent scaling. The measured frequency dependence is then represented and validated against the P.840 prediction.

## II. MEASUREMENT CONFIGURATION AND ATTENUATION-LWC REGRESSION

### *A. Measurement Configuration*

The experimental platform is shown in Fig. 1(a). The transmitter and receiver were placed near one end of an enclosed 13 × 6 × 3 m$^3$ ($L$ × $W$ × $H$) chamber room, and a rigidly fixed 2 × 1.5 m$^2$ ($L$ × $W$) reflector was positioned 11 m away, folding the channel to a total length $L$ = 22 m. A Ceyear 1465D signal generator drove a Ceyear 82406B ×12 multiplier.

This work was supported by the National Natural Science Foundation of China (62471033), and the Natural Science Foundation of Hebei Province (F2026105021). (*Corresponding: jianjun_ma@bit.edu.cn*)
Jiabiao Zhao, Xiaoxiang Li, Kefeng Huang, Yapeng Ge, Hanchen Liu, Jie Yang, Weidong Hu, Houjun Sun and Jianjun Ma are with the School of Integrated Circuits and Electronics, Beijing Institute of Technology, Beijing 100081, China.

Matched HD-1400SGAH25 horn-lens assemblies (40 dBi antenna-lens gain at 140 GHz) terminated both ends, and the received power was recorded with a Ceyear 71718 power sensor. Detailed views are given in Fig. 1(b)-(d).

The transmitter and receiver were mounted 94 cm above the floor, well above the first-Fresnel-zone radius (7.67 cm at 140 GHz), so residual ground reflection is negligible across the band. Water fog was produced by a commercial ultrasonic generator (rated 30 kg/h) that filled the chamber room with droplets of 1-40 µm diameter. At the highest frequency in this work, 160 GHz, the largest droplets (20 µm radius) give a size parameter $2\pi r/\lambda_0 = 0.067 \ll 1$, placing the entire population firmly in the Rayleigh regime and justifying the P.840 formulation. Injection, mixing, and dissipation produced a quasi-stationary fog environment as shown in Fig. 1(e).

A Furbs FBS-310B laser particle size analyzer measured the DSD over a 2.5 m optical path, and only records within the manufacturer-recommended obscuration range of 3%-60% were retained (see Fig. 1(f) and (g)). The reflector was mechanically secured, and a fan together with reflector heating limited droplet deposition and the associated surface loss. Clear-air measurements set the reference level, whose fluctuation was far smaller than the fog-induced attenuation. Signal generation and power acquisition were coordinated over a LAN-SCPI network with protocol-layer control on an RS-485 bus, and ambient relative humidity and temperature were logged continuously with an Elitech RC-41THE. The run-averaged chamber temperature was T = 20 °C, and ITU-R P.840 is evaluated at this temperature in Section III.

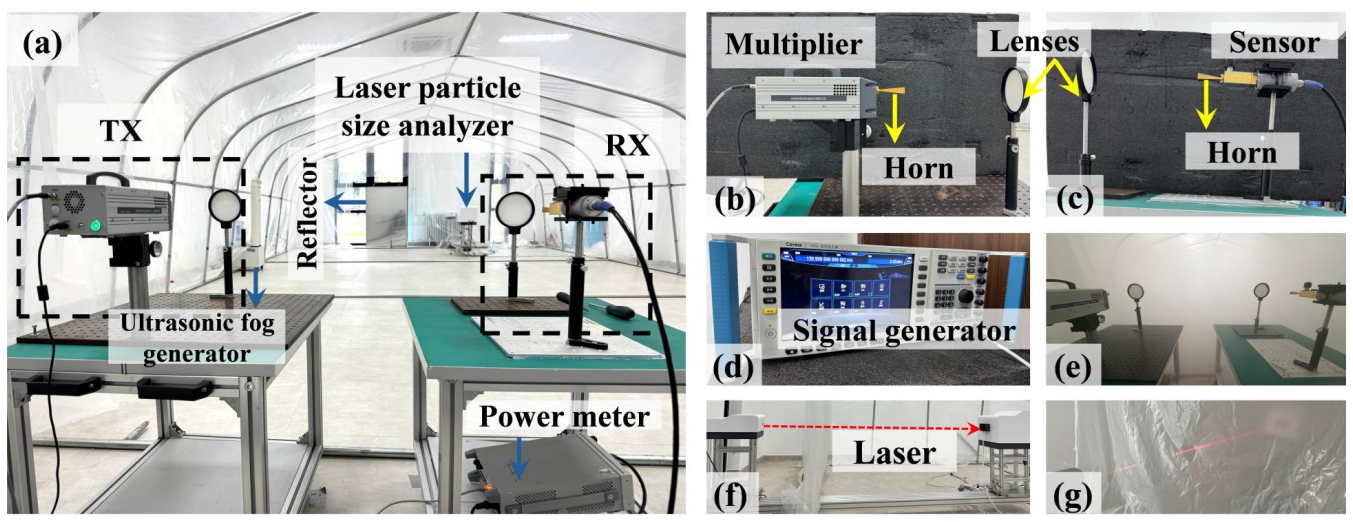

Fig. 1. Experimental setup for channel measurement in foggy room. (a) foggy room with TX/RX, reflector, laser analyzer, fog generator; (b)-(d) multiplier/horn, receiving horn/sensor, signal generator; (e)-(g) fog states and laser path.

### B. LWC Retrieval from the Measured DSD

The analyzer reports the total droplet number concentration together with the volume fraction in discrete diameter bins. The bin-wise volume fractions were converted to number fractions through a third-to-zeroth moment transformation, yielding the per-bin number concentration $N_i(t)$ [15][16]. Assuming spherical droplets, the time-dependent LWC is

$$M(t) = \frac{\pi \rho_w}{6} \sum_{i=1}^{N} d_i^3 N_i(t) \tag{1}$$

where $\rho_w = 1$ g/cm$^3$ is the density of pure water, $d_i$ is the representative diameter of the $i$ th bin, and $t$ denotes time. The 2.5 m optical sampling path runs alongside the THz channel rather than along it and is much shorter than the 22 m propagation path. Because the enclosed, continuously mixed room promotes gradual, quasi-stationary evolution, the sampled DSD is a practical representation of the fog, yet it remains a representative estimate rather than an exact characterization along the entire channel. Accordingly, $M(t)$ is treated as a local proxy for the path-averaged LWC rather than assumed identical to it, and the resulting spatial-representativeness uncertainty is addressed explicitly through the errors-in-variables analysis in Section III.

### C. Fog-Induced Attenuation and Measured Path Slope Extraction

The measured total attenuation ($A_{\text{total}}$) comprises the fog-induced attenuation ($A_f$) and the incremental water-vapor absorption ($A_m$) produced by the humidity rise that accompanies fog generation:

$$A_{\text{total}}(t) = A_f(t) + A_m(t) \tag{2}$$

The water vapor term $A_m(t)$ was computed line-by-line from ITU-R P.676-13 [17][18] using the logged humidity and temperature and subtracted, so that $A_f(t)$ reflects suspended droplets alone. Over the sampled runs this correction is a small fraction of $A_f$ and does not affect the fitted slope beyond its stated uncertainty. Writing the ITU relation on the folded path with $k_{\text{ITU,path}} = K_l \cdot L$, the fog-induced attenuation is

$$A_f(t) = K_l M(t) L \tag{3}$$

For each frequency and run, the time-aligned LWC and $A_f$ samples were fitted by ordinary least squares (OLS) with a free intercept, as

$$A_f(t) = k_{\text{path}} M(t) + b + \varepsilon(t) \tag{4}$$

where $k_{\text{path}}$ is the fitted path-scale attenuation-LWC slope, $b$ is the intercept, and $\varepsilon(t)$ is the residual. For comparison across runs and with P.840, we report the length-normalized specific attenuation $K_{\text{meas}} = k_{\text{path}}/L$ in (dB/km)/(g/m$^3$) as the primary quantity. $k_{\text{path}}$ is retained because it is the quantity directly relevant to a 22 m link budget. Because samples within a run are acquired at second scale while the fog evolves slowly, the residuals $\varepsilon(t)$ are serially correlated and the nominal OLS standard errors are optimistic. Uncertainty is therefore quoted as the dispersion across the three independent runs at each frequency, which is free of within-run autocorrelation. The intercept absorbs reference-level offsets, baseline drift, and the DSD-to-path spatial mismatch, and is not interpreted as physical attenuation at zero LWC.

The locally sampled LWC is an error-contaminated proxy for the path-averaged LWC. Under the classical single-regressor errors-in-variables (EIV) model with additive, mutually uncorrelated errors, the OLS slope is biased toward zero by the reliability ratio of the regressor (regression dilution), and the true slope lies within the reverse-regression (Gini-Frisch) interval [$k_{\text{path}}$, $k_{\text{path}}/R^2$], whose upper end is the reciprocal of the slope of the reverse (LWC-on-attenuation) fit [19][20]. These bounds are used in Section III-B.

## III. Results and Discussion

### A. Linearity and Measured Specific Attenuation

Fig. 2 shows density-colored scatter of $A_f$ against the time-aligned LWC for the nine runs, together with the OLS fits.

Absolute attenuation is not comparable across frequencies, because each run is an independently generated fog realization spanning a different LWC range and temporal evolution. Within every run, attenuation increases with LWC along a near-linear trend, and the high-density clusters indicate that the fog dwelt in several quasi-stationary concentration states rather than sweeping LWC uniformly. The localized departures near cluster transitions reflect temporal evolution, spatial non-uniformity, and the imperfect correspondence between the locally sampled LWC and the path-integrated attenuation.

The sampled LWC extends to roughly 8 g/m$^3$, an order of magnitude beyond natural fog (~ 0.05 g/m$^3$ for medium fog with visibility of order 300 m and 0.5 g/m$^3$ for thick fog with visibility of order 50 m [7]). This wide range increases the regression leverage. Because Rayleigh absorption ties the droplet attenuation to the third moment of the DSD alone, the fitted slope remains transferable to natural fog at the same temperature, so the elevated LWC improves estimation precision rather than restricting applicability.

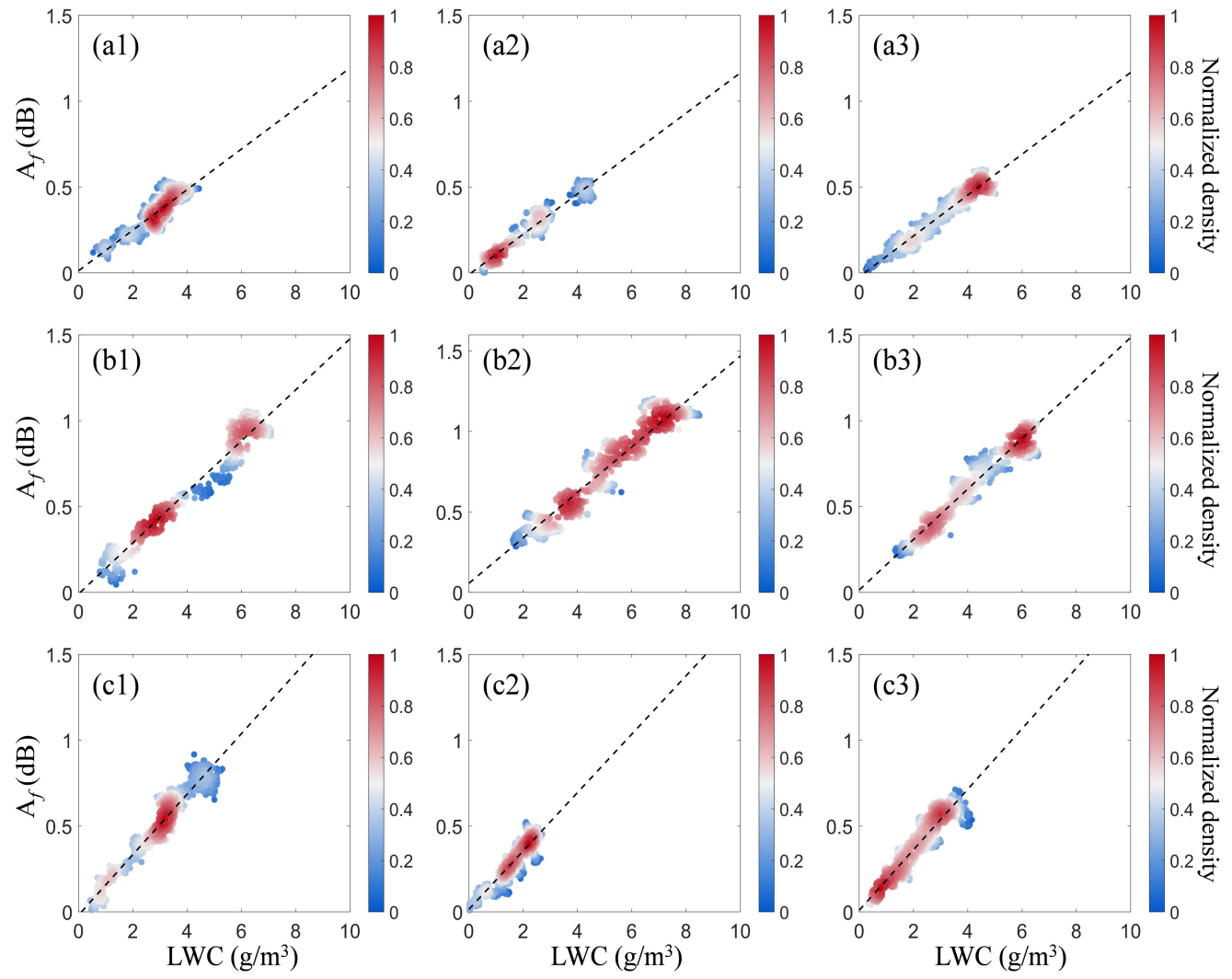

Fig. 2. Density-colored scatter of $A_f$ vs LWC for the three runs at (a1)-(a3) 120 GHz, (b1)-(b3) 140 GHz, (c1)-(c3) 160 GHz, with ordinary least-squares fits (dashed lines).

The per-run regression results are listed in Table I. The fitted $k_{path}$ span 0.1175-0.1191, 0.1412-0.1486, and 0.1701-0.1763 dB/(g/m$^3$) at 120, 140, and 160 GHz, with means 0.1182, 0.1456, and 0.1742 dB/(g/m$^3$) and within-frequency coefficients of variation of 0.68%, 2.67%, and 2.04%. Equivalently, the mean specific attenuation $K_{meas}$ rises from 5.374 to 6.618 to 7.918 (dB/km)/(g/m$^3$) as shown in Table II, a 47.3% increase from 120 to 160 GHz. The coefficients of determination ($R^2$ = 0.799-0.952) and RMSE (0.0381-0.0733) dB support linearity over the sampled ranges while exposing run-dependent residual dispersion. The intercepts span -0.0231 to 0.0565 dB. Run 140/2 is diagnostic, pairing the largest positive intercept (0.0565 dB) with the lowest slope among the 140 GHz runs and the largest ITU shortfall (see Table I) - precisely the joint signature expected when an OLS line pivots through the data centroid under maximal regressor noise [21].

TABLE I. Regression Results for the Nine Fog Measurement Runs

| $f$ / run (GHz) | $k_{path}$ [dB/(g/m$^3$)] | $b$ [dB] | RMSE [dB] | $R^2$ | Frisch upper bound [dB/(g/m$^3$)] | $\Delta_{ITU}$ [%] |
|---|---|---|---|---|---|---|
| 120/1 | 0.1181 | 0.0128 | 0.0496 | 0.799 | 0.1478 | 2.25 |
| 120/2 | 0.1175 | -0.0097 | 0.0386 | 0.930 | 0.1263 | 2.75 |
| 120/3 | 0.1191 | -0.0231 | 0.0381 | 0.944 | 0.1262 | 1.43 |
| 140/1 | 0.1486 | -0.0086 | 0.0648 | 0.952 | 0.1561 | 0.77 |
| 140/2 | 0.1412 | 0.0565 | 0.0733 | 0.923 | 0.1530 | 5.71 |
| 140/3 | 0.1470 | 0.0142 | 0.0550 | 0.946 | 0.1554 | 1.84 |
| 160/1 | 0.1762 | -0.0188 | 0.0562 | 0.942 | 0.1870 | 0.91 |
| 160/2 | 0.1701 | 0.0147 | 0.0457 | 0.875 | 0.1944 | 4.34 |
| 160/3 | 0.1763 | 0.0093 | 0.0576 | 0.902 | 0.1955 | 0.86 |

Note: RMSE and $R^2$ are the regression root-mean-square error and coefficient of determination. The Gini-Frisch (reverse-regression) upper bound is $k_{path}/R^2$. $\Delta_{ITU} = (k_{ITU,path} - k_{path})/k_{ITU,path} \times 100\%$ with $k_{ITU,path} = K_l \cdot L$ evaluated at 20 °C.

*B. Error-Aware Comparison with ITU-R P.840*

For a comparison at the measured propagation length, the P.840 coefficients at 20 °C - recomputed here from the double-Debye water model as $K_l$ = 5.492, 6.807, and 8.083 (dB/km)/(g/m$^3$) at 120, 140, and 160 GHz - were multiplied by $L$ = 0.022 km to give $k_{ITU,path}$. Both $k_{path}$ and $k_{ITU,path}$ increase monotonically across the band as shown in Fig. 3. The individual deviations $\Delta_{ITU}$ range from 0.77% to 5.71%, with run-mean values 2.14%, 2.77%, and 2.04% (see Table II). And the mean $k_{path}$ lie below $k_{ITU,path}$ at all three frequencies.

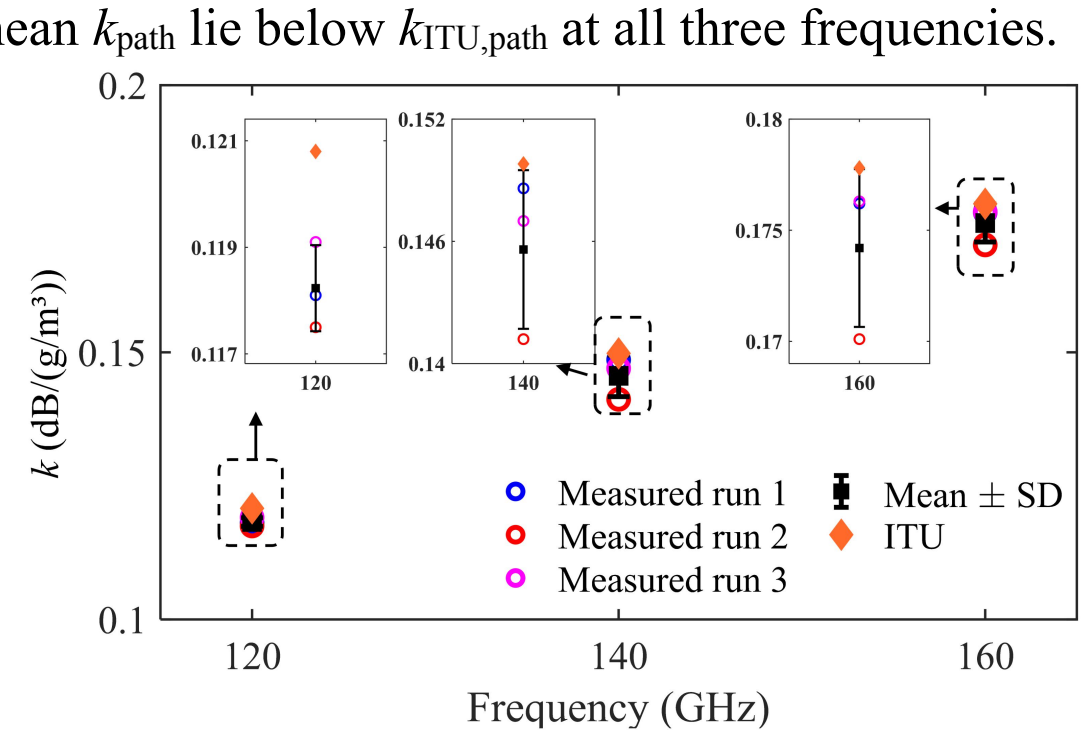

Fig. 3. Measured attenuation-LWC regression slopes ($k_{path}$) vs ITU-predicted $k_{ITU,path}$ at 120/140/160 GHz. Open circles = runs, squares ± SD = mean, diamonds = $k_{ITU,path}$. Insets enlarge the dashed regions.

Two features of this shortfall matter. First, it is one-signed: every one of the nine runs falls below P.840. Random calibration error would scatter in both directions, whereas a strictly one-signed deficit is the signature of attenuation bias, under which the OLS slope of a mismeasured regressor is biased toward zero. Consistent with this, the Gini-Frisch interval [$k_{path}$, $k_{path}/R^2$] contains $k_{ITU,path}$ for all nine runs (see Table I). The bound is informative only where $R^2$ is high. However, its fractional width is $(1 - R^2)/R^2$, about 5% for the best runs ($R^2 \approx 0.95$) but about 25% for run 120/1 ($R^2$ = 0.799), where containment is nearly vacuous. Containment is therefore corroborating rather than decisive, and we do not rest the conclusion on it.

Second, interpreting $k_{ITU,path}$ as the true slope yields an estimated reliability ratio $\widehat{\lambda} = K_{meas}/K_{ITU}$ = 0.979, 0.972, and 0.980 at the three frequencies (see Table II). Its near-constancy is itself evidence. An RF-side calibration error would generally be frequency-dependent, whereas the LWC-proxy error - rooted in the 2.5 m-versus-22 m spatial mismatch and in the DSD retrieval - is a property of the microphysical measurement and hence common to all three frequencies. The implied error-to-signal variance ratio, $(1 - \widehat{\lambda})/\widehat{\lambda} \approx 0.02$-$0.03$, is small and stable across frequency, consistent with a single regressor-side mechanism of modest magnitude.

TABLE II. MEAN MEASURED AND ITU-R P.840 SPECIFIC ATTENUATION COEFFICIENTS, RELIABILITY RATIOS, AND DEVIATIONS

| Frequency (GHz) | $K_{meas}$ (mean) [(dB/km)/(g/m$^3$)] | $K_{ITU}$ (20 °C) [(dB/km)/(g/m$^3$)] | $\widehat{\lambda} = K_{meas}/K_{ITU}$ | Mean $\Delta_{ITU}$ (%) |
|---|---|---|---|---|
| 120 | 5.374 | 5.492 | 0.9786 | 2.14 |
| 140 | 6.618 | 6.807 | 0.9723 | 2.77 |
| 160 | 7.918 | 8.083 | 0.9796 | 2.04 |

*C. Frequency Scaling Law and Temperature Diagnostic*

The P.840 frequency dependence can be validated without resolving the absolute bias. Under the multiplicative model $k_{path}(f) = \lambda \cdot K_l(f) \cdot L$, with a frequency-common factor λ - arising either from regression dilution or from an overall calibration offset - the frequency dependence is carried entirely by the power-law exponent

$$n = \frac{\ln[K(f_2)/K(f_1)]}{\ln(f_2/f_1)} \quad (5)$$

in which $\lambda$ cancels. Here $K$ denotes $K_{meas}$ for the measured coefficients and $K_l$ for the P.840 coefficients. Since P.840 supplies $K_l$ only through the double-Debye model and, unlike rain [8], is not expressed as a power law, this exponent serves simultaneously as a compact engineering scaling law and as a bias-immune benchmark.

The measured exponent is $n$ = 1.347 ± 0.043 over the 120-160 GHz endpoints. 1.347 from a three-point log-log fit, with 1.351 and 1.343 for the two adjacent frequency pairs (120-140 and 140-160 GHz), the between-run dispersion setting the uncertainty. P.840 gives 1.343 at 20 °C, with corresponding pair exponents of 1.392 and 1.286. The measured pairs reproduce the sign of this curvature, the exponent decreasing from the lower to the upper pair. The band-average agreement - 1.347 versus 1.343, or 0.3% - is an order of magnitude tighter than the absolute-coefficient deviations, yet it is obtained with a statistic immune to the common bias of Section III-B. Equivalently, the interfrequency ratios match P.840 to within 0.8% (see Table III). For link budgeting, $K_{meas}(f) \approx 6.62 \cdot (f/140\ \text{GHz})^{1.35}$ (dB/km)/(g/m$^3$) across the band.

The exponent also fixes the temperature. The sensitivity of $K_l$ is frequency-dependent - -0.83, -0.55, and -0.30 %/°C at 120, 140, and 160 GHz - so $K_l$ decreases with warming, most rapidly at the low end. The P.840 exponent rises with temperature at $dn/dT$=+0.018 /°C ($n$=1.247, 1.343, 1.432 at 15, 20, 25 °C), which localizes the effective droplet temperature to 20.2 ± 2.3 °C. This validates the 20 °C reference directly from the data and bounds what temperature alone can explain. At the 1σ edge (+2.3 °C), $K_l$ falls by only 2.0%, 1.4%, and 0.8%, absorbing most of the 120 GHz shortfall but not all three deficits at once, and in a frequency-dependent pattern unlike the frequency-flat reliability ratio observed. Temperature re-referencing and LWC-proxy bias are therefore jointly bracketed - the flat $\widehat{\lambda}$ - favoring the latter - with the logged chamber temperature (Section II-A) supplying the constraint that closes the loop. The scaling is accordingly exact to 0.3% in the exponent, while the one-signed 2.0-2.8% absolute deficit stays within the reverse-regression bounds for all nine runs.

TABLE III. MEASURED AND ITU-R P.840 INTERFREQUENCY RATIOS AND POWER-LAW EXPONENTS AT 20 °C AND 25 °C.

| Quantity | Measured (mean) | P.840, 20 °C | P.840, 25 °C | \|Δ\| (%) |
|---|---|---|---|---|
| 140/120 ratio | 1.2315 | 1.2394 | 1.2562 | 0.64 |
| 160/140 ratio | 1.1964 | 1.1874 | 1.2017 | 0.76 |
| 160/120 ratio | 1.4734 | 1.4717 | 1.5096 | 0.11 |
| $n$ (120-160 GHz) | 1.347 | 1.343 | 1.432 | 0.28 |

Note: |Δ| denotes the absolute percentage deviation of the measured value from the ITU-R P.840 prediction at 20 °C.

## IV. CONCLUSION

This letter characterizes LWC-referenced fog attenuation in the low-THz band. Controlled measurements at 120, 140, and 160 GHz over a 22 m channel were paired with LWC retrieved from a time-aligned DSD. These give the attenuation-LWC relationship at each frequency, with mean specific attenuation coefficients of 5.37, 6.62, and 7.92 (dB/km)/(g/m$^3$) that are consistent with ITU-R P.840. Such reference data were not previously available at these frequencies, where earlier studies relied on visibility. Casting the P.840 comparison as an errors-in-variables problem separates the absolute coefficient from its frequency dependence and yields a compact power-law scaling law. Because it depends only on interfrequency ratios, the measured exponent of 1.347, matching the P.840-implied 1.343, is insensitive to any scaling shared across frequencies and provides a robust benchmark for the frequency dependence. The exponent further constrains the effective droplet temperature to near 20 °C, consistent with the chamber record. These results provide reference data and a compact scaling law for low-THz fog link budgeting, and they indicate the potential of multifrequency links for opportunistic fog-microphysics sensing.